\documentclass[floats,floatfix,showpacs,amssymb,prl,twocolumn,superscriptaddress,nofootinbib]{revtex4-2}
\usepackage[dvipsnames]{xcolor} 
\usepackage{amsmath} 
\usepackage{booktabs}
\usepackage[utf8]{inputenc}
\usepackage{graphicx} 
\usepackage[colorlinks=true,citecolor=blue,urlcolor=blue]{hyperref}
\usepackage{lineno}
\usepackage[normalem]{ulem}
\usepackage{xspace}

\newcommand{\ssout}[1]{}

\newcommand{\beq}{\begin{equation}}
\newcommand{\eeq}{\end{equation}}
\newcommand{\beqa}{\begin{eqnarray}}
\newcommand{\eeqa}{\end{eqnarray}}

\newcommand{\msun}{\ensuremath{M_{\odot}}\xspace}

\newcommand{\paper}{{\it Letter}\xspace}
\newcommand\prlsec[1]{\vspace{2mm}\noindent \emph{#1}--}
\newcommand{\catalog}{GWTC-3\xspace}
\newcommand{\tjn}{\ensuremath{\theta_{\rm{JN}}}\xspace}
\newcommand{\atjn}{\ensuremath{\alpha_{\tjn}}\xspace}
\newcommand{\btjn}{\ensuremath{\beta_{\tjn}}\xspace}

\newcommand{\nn}{\nonumber}

\newcommand{\vl}{\ensuremath{\vec{\lambda}}\xspace}
\newcommand{\vd}{\ensuremath{D}\xspace}

\newcommand{\vt}{\ensuremath{\vec{\theta}}\xspace}
\newcommand{\ud}{\ensuremath{\mathrm{d}}}
\newcommand{\maHP}{\ensuremath{\mathcal{H}_{\mathrm{PE}}}\xspace}

\begin{document}
\title{The orientations of the binary black holes in \catalog}
\author{Salvatore Vitale}
\email{salvatore.vitale@ligo.org}
\author{Sylvia Biscoveanu}
\author{Colm Talbot}
\affiliation{LIGO, Massachusetts Institute of Technology, 77 Massachusetts Avenue, Cambridge, MA 02139, USA}
\affiliation{Kavli Institute for Astrophysics and Space Research and Department of Physics, Massachusetts Institute of Technology, 77 Massachusetts Avenue, Cambridge, MA 02139, USA}
\date{\today}

\begin{abstract}
It is expected that the orbital planes of gravitational-wave (GW) sources are isotropically distributed. However, both physical and technical factors, such as alternate theories of gravity with birefringence, catalog contamination, and search algorithm limitations, could result in inferring a non-isotropic distribution.
Showing that the inferred astrophysical distribution of the orbital orientations is indeed isotropic can thus be used to rule out some violations of general relativity, as a null test about the purity of the GW catalog sample, and as a check that selection effects are being properly accounted for. We augment the default mass/spins/redshift model used by the LIGO-Virgo-KAGRA Collaboration in their most recent analysis to also measure the astrophysical distribution of orbital orientations. We show that the 69 binary black holes in \catalog are consistent with having random orbital orientations. The inferred distribution is highly symmetric around $\pi/2$, with skewness $\mathcal{S}_{\rm{post}}=0.01^{+0.17}_{-0.17}$.  Meanwhile, the median of the inferred distribution has a Jensen–Shannon divergence of $1.4\times 10^{-4}$ bits when compared to the expected isotropic distribution. 
\end{abstract}

\maketitle
\prlsec{Introduction} The latest gravitational-wave (GW) catalog released by the LIGO~\cite{TheLIGOScientific:2014jea}-Virgo~\cite{TheVirgo:2014hva}-KAGRA~\cite{Somiya:2011np, Aso:2013eba, Akutsu:2017kpk} collaboration (LVK) -- GWTC-3~\cite{LIGOScientific:2021djp} -- contains 69 binary black holes (BBH) with false alarm rate smaller than 1 per year~\cite{LIGOScientific:2021psn}. Groups not affiliated with the LVK have reported the discovery of other BBHs, usually with low signal-to-noise ratios (SNRs)~\cite{Nitz:2021zwj,Olsen:2022pin}.
The growing BBH dataset has been used to study the underlying distribution of masses and spins, as well as the evolution of their merger rate with redshift. 
While the primary masses of the BBHs in \catalog are still consistent with a rather simple phenomenological model (a power law plus a gaussian distribution~\cite{Talbot:2018cva}), there is now tentative evidence for extra structure at around 10~\msun and 17~\msun~\cite{Tiwari:2020otp, Edelman:2021zkw, Li:2021ukd, Veske:2021qis, LIGOScientific:2021psn}. Meanwhile, no evidence has been found for either a lower (i.e. in the mass range between $2$~\msun and $\sim 5~\msun$~\cite{2010ApJ...725.1918O,2011ApJ...741..103F}) or an upper mass (i.e. above $\sim 50~\msun$, in the pair instability mass gap~\cite{Fryer:2000my}) gap~\cite{Farah:2021qom}. The distribution of the spin magnitudes peaks at around $0.2$, if both spins are treated as identically distributed~\cite{Wysocki:2018, LIGOScientific:2021psn}, while that of the most rapidly spinning black hole in each binary peaks at $\sim 0.4$~\cite{Biscoveanu:2020are}. This results in a distribution for the effective inspiral spin $\chi_{\rm{eff}}$ that favors small ($\sim 0.1$) positive values~\cite{Miller:2020zox, LIGOScientific:2021psn}. In fact, what can be learned from the spins of \catalog is still debated in the literature~\cite{Roulet:2021hcu,Galaudage:2021rkt,Miller:2020zox}, and more will be learned in the fourth observing run, starting in late 2022. Ref.~\cite{Callister:2021fpo} has reported evidence for an anti-correlation between the mass ratio and the effective inspiral spin of the BBHs in the {44} BBHs of the previous LVK catalog, GWTC-2~\cite{LIGOScientific:2020ibl}. This finding has later been confirmed by the LVK in \catalog, at a similar level of significance~\cite{LIGOScientific:2021psn}. Recent works have also found evidence of less significant correlations between the other binary mass parameters and the effective spin distribution~\cite{Safarzadeh:2020mlb, Franciolini:2022iaa,2021arXiv211013542H}.
While measurements described above pertain to the intrinsic parameters of the sources, the extrinsic source parameters can also be used to learn about the BBH population. The LVK has shown that the BBH merger rate evolves with redshift in a way that is consistent with the star formation rate~\cite{LIGOScientific:2021psn}. Estimates of source distances can also be used to exclude alternatives to general relativity~\cite{Pardo:2018ipy} and to measure the speed of gravitational waves~\cite{LIGOScientific:2017zic}.

In this \paper we measure the astrophysical distribution of another important extrinsic parameter, the orbital inclination \tjn, defined as the angle between the total angular momentum and the line of sight. It might seem obvious that the orientation of BBH orbits should be isotropic, i.e. that $\pi(\tjn) \propto \sin\tjn$~\cite{Maggiore:2007ulw,Schutz:2011tw}. However, several factors may result in detecting a distribution for \tjn which is not isotropic. First, some alternative theories of gravity, e.g. birefringence, predict that the GW power emitted by a merging binary is not symmetric about the orbital plane~\cite{Alexander:2009tp,Zhao:2019xmm,Okounkova:2021xjv}: this would result in a preference for detecting sources with inclination of e.g. $\pi/6$ over sources with inclination of $\pi - \pi/6$, whereas these two values would be equally detectable in general relativity. 
Second, the algorithms that search for compact binary coalescences (CBCs) assume that spins are perfectly aligned with the orbital angular momentum, {and neglect higher order modes (HOMs)}~\cite{LIGOScientific:2021djp}. These approximations have been shown to be sufficient when spins are small or aligned with the orbit, and mass ratios are close to unity. However, their accuracy is also a function of the orbital inclination: indeed, the loss of SNR due to neglecting these effects is maximal when $\tjn = \pi/2$ (``edge-on'' orientation). This could result in systematically missing sources with high orbital inclination and hence in a dip in the inferred \tjn distribution at $\tjn = \pi/2$. If this loss of SNR is properly accounted for when calculating selection effects, one should still measure an isotropic distribution. 
Finally, marginal triggers of non-astrophysical origin (often called glitches) could be mistakenly identified as BBHs, and added in the catalog. It is has been found that that running CBC source characterization algorithms on glitches results in distributions of \tjn peaked at $\pi/2$.
A sub-population of glitches in the catalog would thus result in an excess of posterior probability at around $\pi/2$ for \tjn.
Measuring a distribution of \tjn different from the expected isotropic one could thus indicate problems with our understanding of gravity, with the sensitivity of detection algorithms to highly inclined orbits, or with glitch contamination. Ref.~\cite{Okounkova:2021xjv}  used the sources in GWTC-2 to recast the measured level of symmetry of the inclination angles around $\pi/2$ into constraints on birefringence and Chern-Simons gravity. They did not perform full hierarchical inference on the astrophysical distribution of inclination angles, which is the focus of this work.

\prlsec{Method} We parametrize the distribution of the masses, spins, redshifts and inclination angles of BBHs in terms of vector of (unknown) hyper-parameters \vl (``hyper'' to distinguish them from the parameters of individual sources, e.g. masses and spins). Our goal is to infer \vl given the dataset \vd consisting of the 69 \catalog BBHs with false alarm ratio smaller than 1 per year~\cite{LIGOScientific:2021psn}, $\vd \equiv  \{d_i, i=1\ldots 69\}$. If one is not interested in measuring it, the overall merger rate density can be analytically marginalized over to obtain~\cite{Mandel:2018mve,Fishbach:2018edt,Vitale:2020aaz}

\beq
p(\vl | \vd) \propto {\pi(\vl)}  \prod_{i=1}^{69} \frac{p(d_i |\vl) }{\alpha(\vl)}\,. \label{Eq.HyperPostOfLikeFirst} \nn
\eeq

In this expression, $\alpha(\vl)$ represents the \emph{fraction} of detectable BBHs given the population parameters \vl; $\pi(\vl)$ is the population prior and $p(d_i |\vl) $ is the likelihood of the stretch of data containing the i-th BBH. 
The likelihood of the individual sources can be written in terms of their posterior distributions by marginalizing over the individual source parameters $\vt\equiv (m_1, m_2, a_1, \tau_1, a_2, \tau_2,z, \tjn)$. In this expression, $m_1$ and $m_2$ are the masses of the heavier and lighter black hole in the binary, respectively;  $a_i$ is the dimensionless spin of the i-th black hole in the binary, and $\tau_i$ is the angle between the i-th spin vector and the orbital angular momentum, 
$z$ is the redshift and \tjn  is the orbital inclinations. For the i-th source, one has~\cite{2019PASA...36...10T,Vitale:2020aaz}:

\begin{equation}
p(d_i |\vl) =\int \ud \vt p(d_i |\vt ) \pi(\vt|\vl) =  \int \ud \vt\;\; \frac{  p(\vt | d_i \maHP)\pi(\vt|\vl) }{\pi(\vt|\maHP)},\nn
\end{equation}
where $p(\vt | d_i \maHP)$ is the posterior distribution for the source in the i-th stretch of data, and \maHP symbolizes all of the settings that went into the parameter estimation algorithm used to produce those samples. Similarly, $\pi(\vt|\maHP)$ are the priors used when sampling the posterior distribution. Finally, $\pi(\vt|\vl)$ is the population prior, i.e., our model for how the hyper parameters affect the true underlying distribution of the BBH parameters \vt. 
We use the posterior samples of the 69 BBHs reported in \catalog, as released by the LVK in Refs~\cite{GWTC1Release,GWTC2Release,ligo_scientific_collaboration_and_virgo_2021_5117703,ligo_scientific_collaboration_and_virgo_2021_5546663} and approximate the integral with a discrete sum 

\beq
\int \ud \vt p(d_i |\vt ) \pi(\vt|\vl) \simeq {N_{\rm{samples}}}^{-1} \sum_{k}^{N_{\rm{samples}}}\frac{\pi(\vt^k_i|\vl) }{\pi(\vt^k_i|\maHP)} \nn
\eeq
where the ${N_{\rm{samples}}}$ samples are drawn from the posterior distribution of the i-th event. We sample the hyper posterior with the \texttt{GWPopulation} algorithm~\cite{2019PhRvD.100d3030T}, using the \textsc{dynesty}~\cite{Speagle_2020} sampler. \texttt{GWPopulation} requires that the same number of ${N_{\rm{samples}}}$ is used for all sources. We are thus limited by the source for which fewest samples were made public. That is GW200129\_065458, for which ${N_{\rm{samples}}}=3194$.
For the sources reported in GWTC-1, we use the samples labelled \texttt{IMRPhenomPv2\_posterior} in the data release; for GWTC-2 we use \texttt{PublicationSamples}; for GWTC-2.1 we use \texttt{PrecessingSpinIMRHM}, and for GWTC-3 we use \texttt{C01:Mixed}.

The detection efficiency $\alpha(\vl)$ can also be calculated through an approximated sum starting from a large collection of simulated BBHs for which the SNR (or other detection statistic) is recorded, as described in Ref.~\cite{2019RNAAS...3...66F,LIGOScientific:2021psn}. We use the \texttt{endo3\_bbhpop-LIGO-T2100113-v12-\\1238166018-15843600.hdf5}  sensitivity file released by the LVK~\cite{ligo_scientific_collaboration_and_virgo_2021_5546676} to calculate $\alpha(\lambda)$, using a false alarm threshold of 1 per year to identify detectable sources, consistently with~\cite{LIGOScientific:2021psn}. We stress that this file only contains BBHs which would be detectable by GW observatories at O3 sensitivity, whereas our dataset also includes sources detected in O1 and O2. Unfortunately, the sensitivity files released by the LVK in Ref.~\cite{ligo_scientific_collaboration_and_virgo_2021_5636816}  which also cover GWTC-1 and GWTC-2 do not include the inclination of the simulated sources, and thus cannot be used in this work. We have generated our own sensitivity files which account for the changing sensitivity of the GW network since O1 and verified that the results presented below don't change significantly. This makes sense because it is only with next-generation detectors that one should expect a different distribution of \emph{detectable} inclinations~\cite{Schutz:2011tw,Vitale:2016aso} and since O3 does dominate the overall surveyed time-volume. To make our findings easy to reproduce, we therefore only present results obtained with the LVK data products. Finally, we have verified that using only GW BBHs from O3 one obtain consistent results, though with larger error bars.

\prlsec{Models} The population distribution of masses, spins and redshifts are the same as those used by the LVK in their default analysis (\texttt{Fiducial population mass and redshift} and \texttt{Fiducial population spin} in Ref.~\cite{LIGOScientific:2021psn}). Specifically, the distribution of primary masses $m_1$ is parametrized as the mixture model of a power law and a gaussian distribution (7 parameters: power law slope, mixture fraction, minimum and maximum black hole mass, smoothing factor, mean and standard deviation of the gaussian component)~\cite{Talbot:2018cva}; the mass ratio is parametrized as a power law (1 parameter: power law slope)~\cite{Fishbach:2018edt}; the spin magnitudes are parametrized as identically distributed beta distributions (2 parameters)~\cite{Wysocki:2018}; the spin tilts are modeled as identically distributed mixtures of an isotropic component  and a gaussian component centered at zero degrees (2 parameters: the mixture fraction and the width of the preferentially spin-aligned distribution)~\cite{Talbot:2017yur}. 
We use a beta distribution to parametrize the population distribution of inclination angles, rescaled to take values in the range $[0,\pi]$, and parametrized by two hyper parameters \atjn and \btjn:

\beq
p(\tjn | \alpha_{\tjn}, \beta_{\tjn}) = \frac{x^{\atjn - 1} (\pi - x)^{\btjn - 1}}{B(\atjn, \btjn) \pi^{\atjn + \btjn + 1}}\nn
\eeq
with $B(a,b)\equiv \Gamma(a) \Gamma(b)/\Gamma(a+b)$. For $\atjn=\btjn$ this distribution is symmetric around $\pi/2$: for $\atjn=\btjn\simeq 2.15$ it is very similar to $\sin\tjn/2$ (The Jensen–Shannon (JS) divergence~\cite{61115, kullback1951} between $\sin(\tjn)/2$ and $p(\tjn |2.15,2.15)$ is $\sim 5\times 10^{-5}$ bits.), whereas for $\atjn=\btjn > 2.15$ ($\atjn=\btjn < 2.15$) more (less) posterior weight is placed at $\tjn=\pi/2$

For the mass, spin, and redshift hyper parameters we use the same priors used by the LVK, as described in Table VI and Table XII of Ref.~\cite{LIGOScientific:2021psn}, with the exception of the two hyper parameters of the spin magnitude beta distribution, for which we used $\alpha_{\chi}: \texttt{Uniform(1,5)}$, $\beta_{\chi}: \texttt{Uniform(1,5)}$.  For the inclination model we use the following priors: $\atjn: \texttt{Uniform(1,8)}$, $\btjn: \texttt{Uniform(1,8)}$, where the model can be normalized.

\begin{figure}[t]
\includegraphics[width=0.5\textwidth]{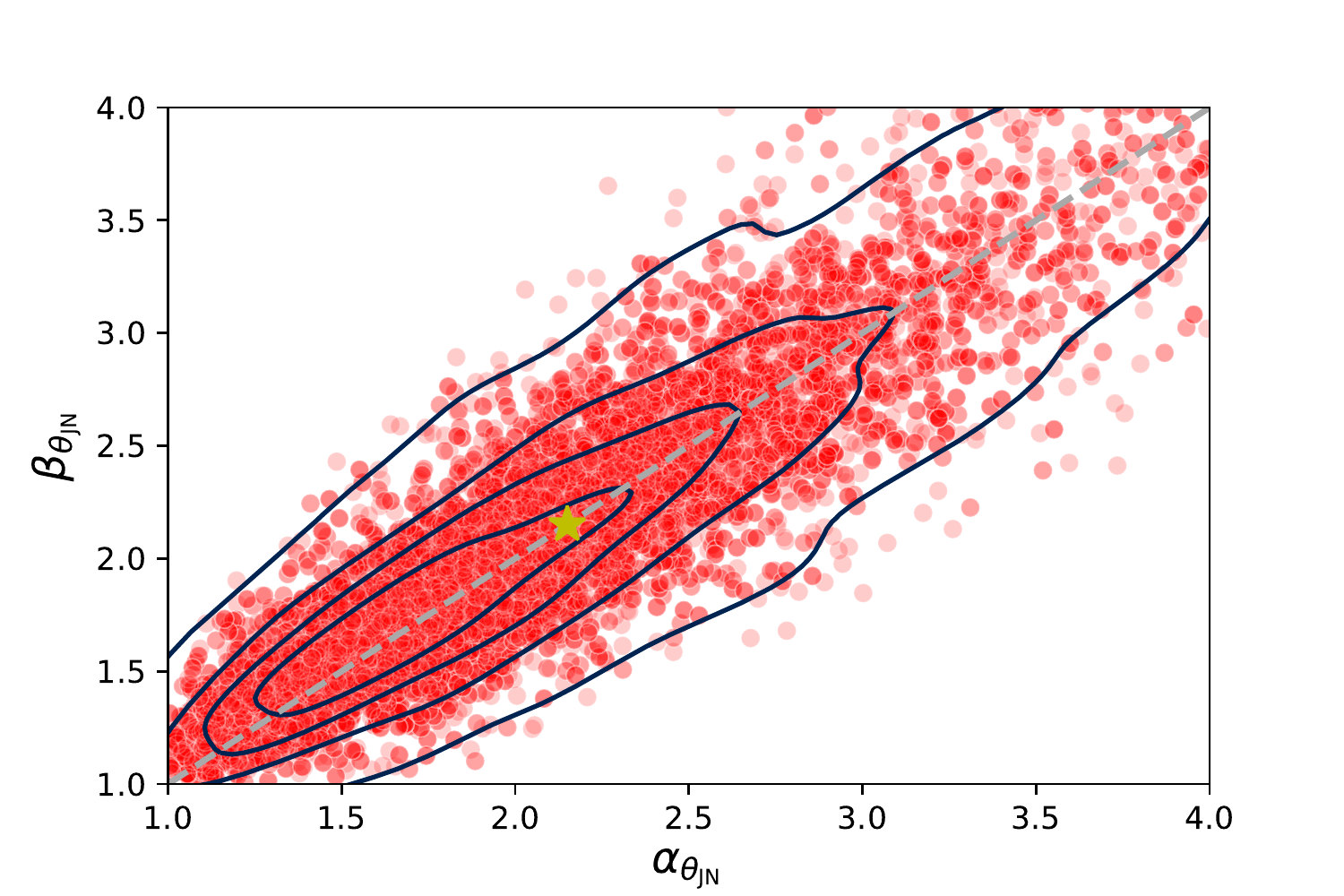}
\caption{Joint distribution for the hyper parameters of the beta model. The posterior lies along the diagonal dashed line, corresponding to distributions which are symmetric around $\tjn=\pi/2$. The point $\atjn=\btjn=2.15$ (marked with a star), corresponding to an approximately isotropic distribution, is included in the top 20\% of posterior mass. }\label{fig.betamarg}
\end{figure}

\prlsec{Results} In Fig.~\ref{fig.betamarg} we show the join distribution of \atjn and \btjn, with KDE contours that increment by steps of 20\% of posterior mass. The point $(\atjn,\btjn) = (2.15,2.15)$ which corresponds to a nearly isotropic distribution is found within the top 20\% of posterior probability. The data yields a joint distribution which is strongly correlated and elongated along the diagonal, indicating a preference for a \tjn distribution which is symmetric around $\pi/2$. The level of symmetry can be be quantified by calculating the skewness of the beta distribution, $\mathcal{S}= 2\frac{\btjn-\atjn}{\btjn+\atjn+2} \sqrt{\frac{\btjn+\atjn+1}{\btjn+\atjn}}$. Drawing 10000 random samples from the hyper posterior, we find $\mathcal{S}_{\rm{post}}=0.01^{+0.17}_{-0.17}$. For comparison, 10000 random samples from the \emph{prior}  yield $\mathcal{S}_{\rm{prior}}=0.00^{+0.88}_{-0.86}$, which shows how the GW data narrows down by a factor of $\sim 5$ the possible skewness of the orbital orientation distribution, relative to the prior we used. 

\begin{figure}[t]
\includegraphics[width=0.5\textwidth]{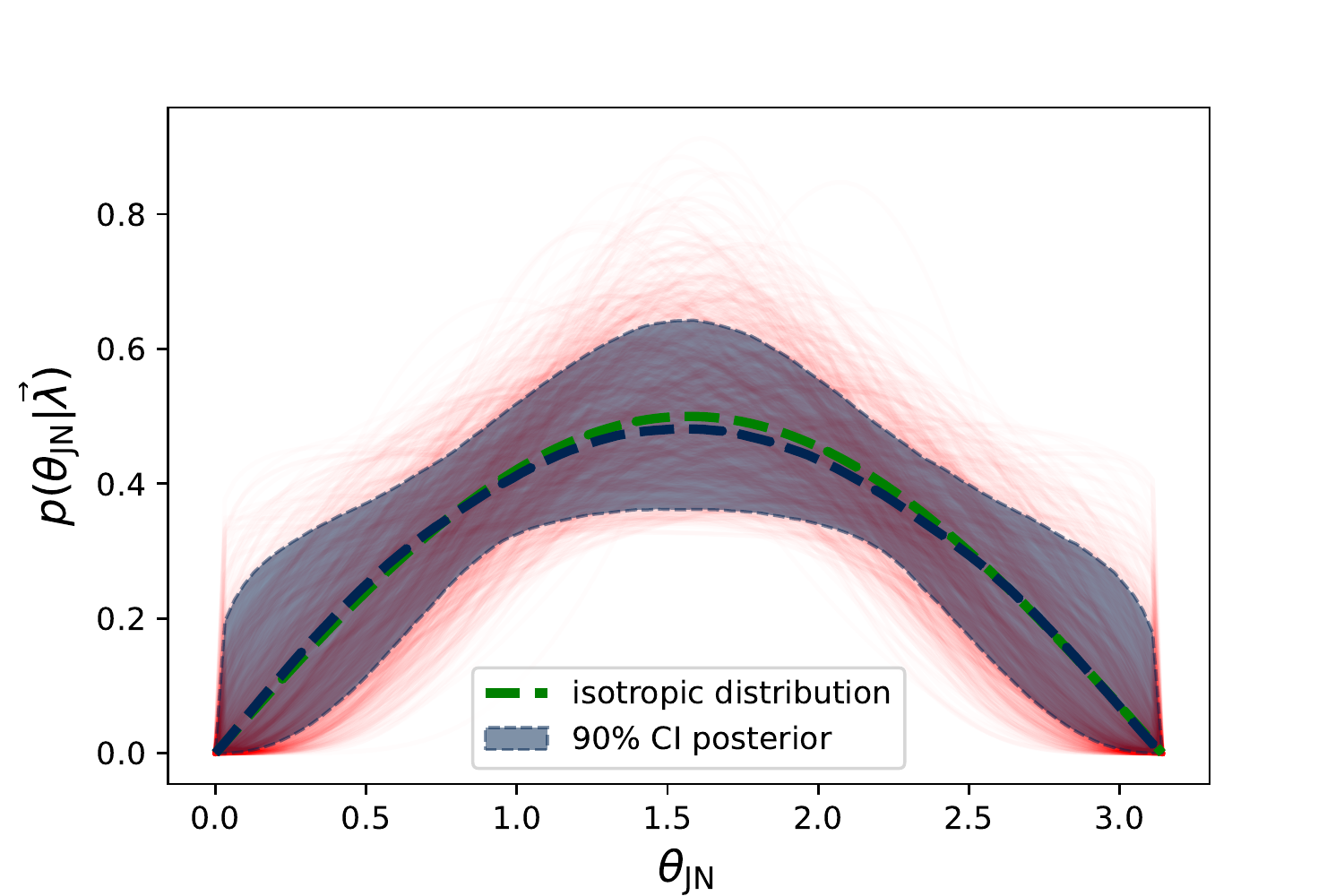}
\caption{Measured population distribution of inclination angles. The dim red lines are 1000 individual draws from the posterior. The blue band shows the 90\% credible interval and blue dashed line its median. Finally, the green dashed line show the expected curve if the source orientations were isotropic. 
}
\label{fig.betapop}
\end{figure}

\begin{figure}[t]
\includegraphics[width=0.5\textwidth]{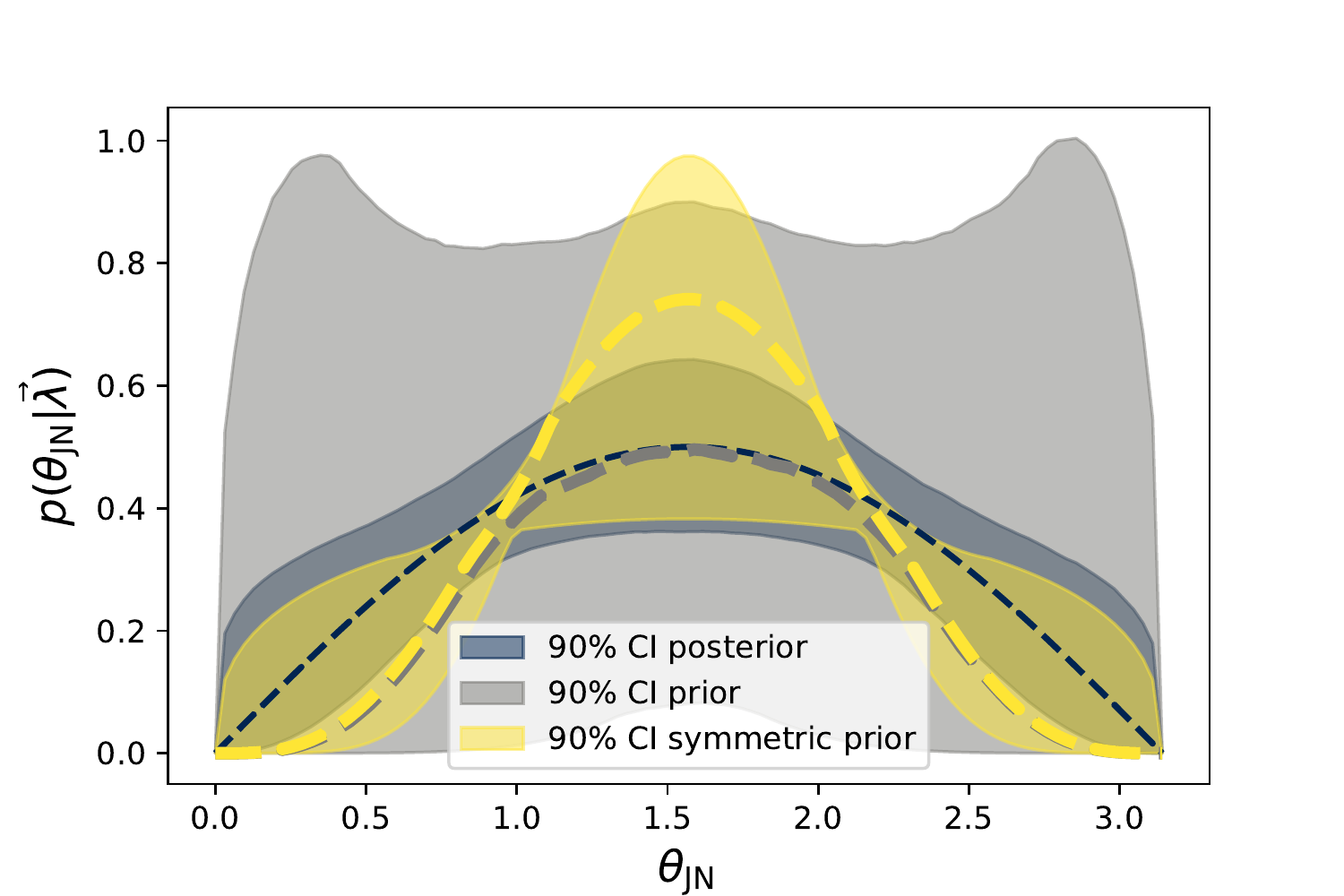}
\caption{90\% credible intervals (colored band) and medians (thick dashed line) for the inclination distribution obtained by sampling the full prior (grey) and a prior that enforces symmetry around $\pi/2$ (yellow). The blue band and dashed line show the 90\% credible interval of the hyper posterior and its median from Fig.~\ref{fig.betapop}, for comparison.}\label{fig.betaprio}
\end{figure}

Next, we take 10000 random samples from the hyper posterior and calculate the corresponding $p(\tjn | \alpha_{\tjn}, \beta_{\tjn})$, Fig.~\ref{fig.betapop}. Each dim red line is an individual draw from the posterior (to enhance visibility, we only show 1000 such draws), whereas the blue band shows the 90\% credible interval. The blue thick curve represents the median, which shows remarkable agreement with the expected isotropic distribution (green dashed curve). The JS divergence between the inferred population median and a perfectly isotropic distribution is $1.4\times 10^{-4}$ bits.
We stress that this level of agreement is not built-in in our model, nor in the shape or width of the hyper parameters' priors. 
This is shown explicitly in Fig.~\ref{fig.betaprio}: first, we draw 10000 points from the \emph{priors} of \atjn and \btjn and plot the resulting 90\% credible interval as a grey band, and the median as a dashed grey curve. The posterior 90\% credible interval and median from Fig.~\ref{fig.betapop} are also shown in blue. It is apparent that the median and the 90\% credible interval are significantly different from the prior. 

However, one might still wonder whether the main information extracted from the data is that the distributions should be symmetric, i.e. that $\atjn \simeq \btjn$. To check whether this is the case, we draw 10000 samples from the prior of $\atjn$ and then calculate $p(\tjn | \alpha_{\tjn}, \beta_{\tjn}=\atjn)$, i.e. we consider the effective prior one would obtain by only considering distributions symmetric around $\pi/2$. The resulting 90\% credible interval is shown as a yellow band in Fig.~\ref{fig.betaprio}, and its median as a dashed yellow curve. Here too it is clear that the measurement is informative, and the data excludes large regions of parameter space. 
The ratio of the JS divergence between the median obtained with prior draws and a perfectly isotropic distribution and the JS divergence calculated with the population posterior median is 288. Using samples from a prior that has been forced to be symmetric, the ratio is 372. 

\prlsec{Conclusion and outlook} The growing set of GW sources can be used to infer the intrinsic properties of the underlying astrophysical populations, but also to learn about cosmology, dark matter, and to perform tests of general relativity. In this \paper we have augmented the main population model used by the LVK in the analysis of \catalog~\cite{LIGOScientific:2021djp} to also infer the underlying distribution of the orbital inclination for the 69 BBHs included in \catalog, which is expected to be isotropic. However several factors can result in inferring an anisotropic distribution: some violations of general relativity; inaccurate evaluation of selection effects due to limitations in the algorithms that search for compact binaries; the inclusion of non-astrophysical triggers (glitches) in the BBH catalog. 
A measurement of the population distribution of orbital orientations can thus be used either as a test of general relativity or, within general relativity, as a null test that all of the analysis components behave as expected. 

We have modeled the astrophysical distribution of orbital inclinations as a non-singular beta distribution, and found that the data prefers inclination distributions which are symmetric around $\pi/2$: the skewness of the inclination distribution is reduced by a factor of 5 relative to what would be obtained using prior samples. Furthermore, the inferred distribution for the inclination angle is consistent with being isotropic: the JS divergence between the inferred population median and a perfectly isotropic distribution is $1.4\times 10^{-4}$ bits. 

We have only used BBH sources reported by the LVK collaboration because a key part of any hierarchical analysis is a self-consistent evaluation of the selection effects, which is hard to achieve for sources found by other groups. However, we do encourage groups that produce independent catalogs of GW sources to also make public their own sensitivity files, to make this type of analysis possible. 
As the size of the GW network increases in the next few years, we expect this analyses will yield even stronger constraints, since a larger network can better reveal the polarization content, and hence the inclination angle, of individual sources. The improved inclination measurement for each source, together with the higher detection rates, implies we could set limits on even smaller departures from orientational isotropy, and use more sophisticated models that explicitly allow for these anisotropies. This will be explored in a future paper.

\prlsec{Acknowledgments} S.V., S.B. and C.T. \ acknowledge support of the National Science Foundation and the LIGO Laboratory. LIGO was constructed by the California Institute of Technology and Massachusetts Institute of Technology with funding from the National Science Foundation and operates under cooperative agreement PHY-0757058.
S.V. is also supported by NSF PHY-2045740.
S.B. is also supported by the NSF Graduate Research Fellowship under Grant No. DGE-1122374.
C.T. is supported by the MIT Kavli Postdoctoral Fellowship.
The authors would like to thank Hsin-Yu Chen, Thomas Dent, Reed Essick, Will Farr, Carl Haster, Charlie Hoy, Ken Ng and Geraint Pratten for insightful discussions and comments.
The authors are grateful for computational resources provided by the LIGO Lab and supported by NSF Grants PHY-0757058 and PHY-0823459.
This paper carries LIGO document number LIGO-P2200104.
\bibliography{draft}

\end{document}